\documentclass[journal]{vgtc}
\onlineid{1793}

\PassOptionsToPackage{svgnames}{xcolor}
\usepackage{xcolor} 
\usepackage{multirow}
\usepackage{array}
\usepackage{url}

\usepackage{tabu}  
\usepackage{tabularx}
\usepackage{amsthm}
\newtheorem{lemma}{Lemma}
\newtheorem{definition}{Definition}
\usepackage{nicefrac}

\usepackage{booktabs}  
\usepackage{wrapfig}
\usepackage{float}

\usepackage{lipsum}                   
\usepackage{mwe}    

\usepackage{algorithm}
\usepackage{algorithmic}
\usepackage{amsfonts}
\usepackage{amsmath}
\usepackage{wrapfig}
\usepackage{siunitx}

\usepackage{mathptmx} 

\definecolor{NavyBlue}{rgb}{0.0, 0.0, 0.5}
\definecolor{BrickRed}{rgb}{0.8, 0.25, 0.33}
\definecolor{ForestGreen}{rgb}{0.13, 0.55, 0.13}

\vgtccategory{Research}

\vgtcpapertype{Data Transformations}

\title{MSz: An Efficient Parallel Algorithm for Correcting Morse-Smale Segmentations in Error-Bounded Lossy Compressors}

\author{
  \authororcid{Yuxiao Li}{0000-0002-8715-5982},
  \authororcid{Xin Liang}{0000-0002-0630-1600},
  \authororcid{Bei Wang}{0000-0002-9240-0700},
  \authororcid{Yongfeng Qiu}{0009-0004-3827-8814},
  \authororcid{Lin Yan}{0000-0001-7017-0329},
  and \authororcid{Hanqi Guo}{0000-0001-7776-1834}
}

\authorfooter{
  \item Yuxiao Li, Yongfeng Qiu, and Hanqi Guo are with The Ohio State University. E-mail:
\{li.14025|qiu.722|guo.2154\}@osu.edu.
 \item Xin Liang is with the University of Kentucky. E-mail: xliang@uky.edu.
 \item Bei Wang is with the University of Utah. E-mail: beiwang@sci.utah.edu.
 \item Lin Yan is with the Iowa State University. E-mail: linyan@iastate.edu.
  
}

\abstract{
  This research explores a novel paradigm for preserving topological segmentations in existing error-bounded lossy compressors. Today's lossy compressors rarely consider preserving topologies such as Morse-Smale complexes, and the discrepancies in topology between original and decompressed datasets could potentially result in erroneous interpretations or even incorrect scientific conclusions. In this paper, we focus on preserving Morse-Smale segmentations in 2D/3D piecewise linear scalar fields, targeting the precise reconstruction of minimum/maximum labels induced by the integral line of each vertex. The key is to derive a series of edits during compression time. These edits are applied to the decompressed data, leading to an accurate reconstruction of segmentations while keeping the error within the prescribed error bound. To this end, we develop a workflow to fix extrema and integral lines alternatively until convergence within finite iterations. We accelerate each workflow component with shared-memory/GPU parallelism to make the performance practical for coupling with compressors. We demonstrate use cases with fluid dynamics, ocean, and cosmology application datasets with a significant acceleration with an NVIDIA A100 GPU.
}


\graphicspath{{figs/}{figures/}{pictures/}{images/}{./}} 
\begin{document}

\firstsection{Introduction} \label{sec:intro}

\maketitle

The rapid advancement of high-performance computing (HPC) technologies has enabled the generation of vast quantities of scientific data, posing significant challenges to scientists regarding data storage, transmission, and visualization.  As such, scientists have recently started to explore compression, especially error-bounded lossy compression, to address the data challenges by ensuring efficient data management and utilization while limiting the amount of distortion introduced by compression. 
The adoption of error-bounded lossy compression techniques, exemplified by algorithms like SZ~\cite{sz, sz3, Tao_2017, Kai2021, Kai2020}, ZFP~\cite{zfp}, and FPZIP~\cite{FPZIP}, offers a pragmatic solution for reducing scientific data. These methodologies facilitate substantial data reduction while maintaining a predefined accuracy threshold, thus ensuring the utility of compressed datasets for immediate analysis and scientific exploration.

\begin{figure}[hbt!]
\centering
    \includegraphics[width=\linewidth]{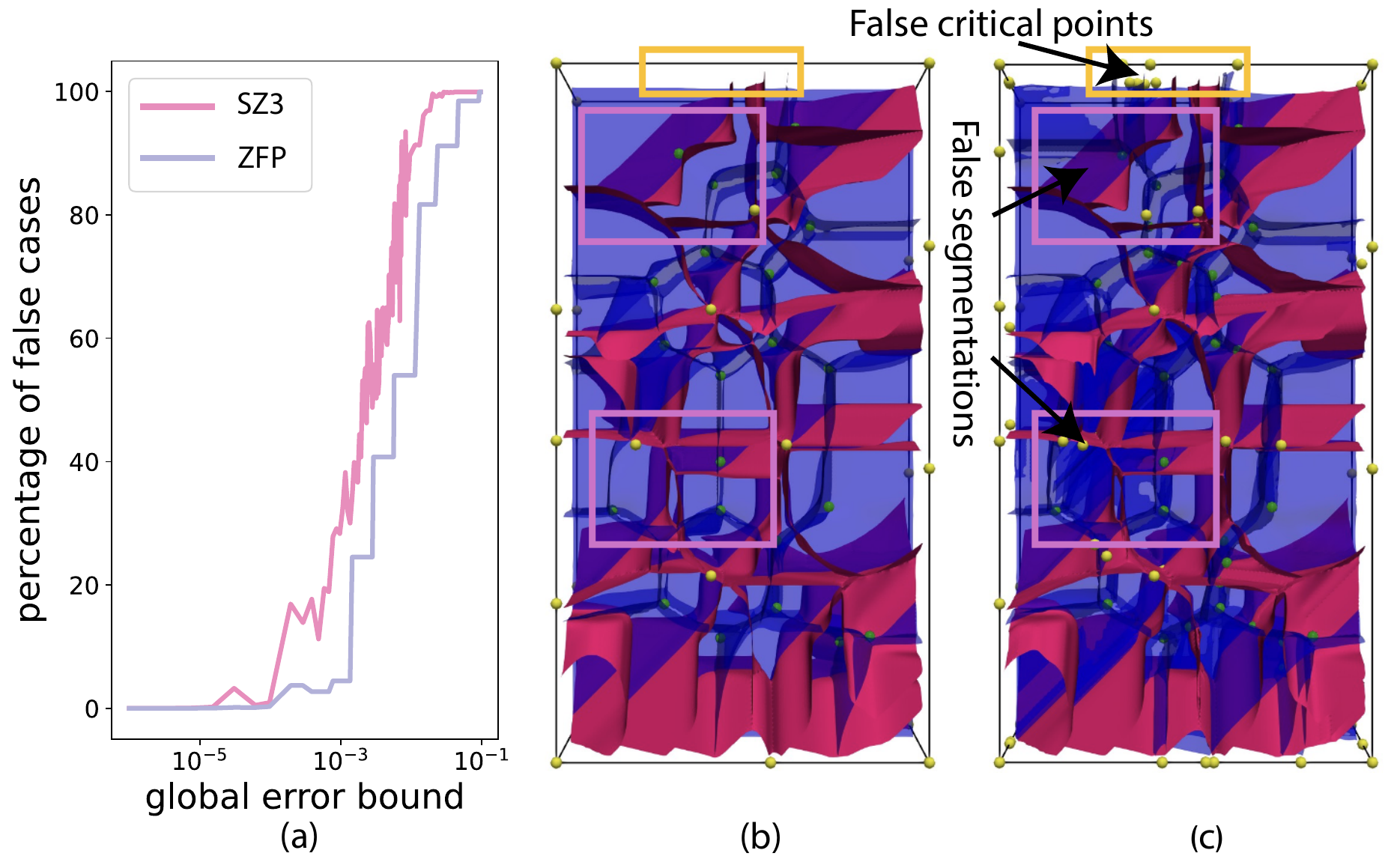}
    \caption{Impacts of lossy compression (SZ3 and ZFP) on MS segmentations of the Adenine Thymine (AT) dataset: (a) percentages of vertices with wrong segmentation labels w.r.t different error bound; (b) and (c): critical points and separatrices in the original data and SZ3's decompressed data with a relative error bound of $10^{-3}$.}
\label{fig:intro} 
\end{figure}

However, an emerging issue with lossy compressors is the inability to preserve topological features such as Morse-Smale (MS) complexes and merge trees in decompressed data, even with bounded error~\cite{yan2023toposz, 9086223}. Topological inconsistencies between the original and decompressed data may result in misinterpretation of the data, even leading to erroneous scientific findings. For example, in molecular dynamics, scientists use MS complexes to segment electron density fields, identifying regions with chemical bonds or weak interactions~\cite{Otero_at}.  Inaccuracies in the segmentations (as exemplified in Figure~\ref{fig:intro}) could lead to wrong interpretations of bonding characteristics and electrostatic interactions. In combustion research, MS complexes help identify reaction, mixing, and quenching zones~\cite{Bremer_combustion, Bremer_flames}.  Erroneous segmentations could lead to flame structure misinterpretations and subsequently affect the analysis of combustion efficiency and reaction mechanisms. In the visualization of Atmospheric Rivers (ARs)~\cite{Lan_ivt}, elongated bands of water vapor transport that originate from the tropics to North America and cause flooding, scientists characterize the skeleton of ARs with MS complexes and segmentations to understand the formation and development of ARs better.  In case studies later in this paper, we will further exemplify how off-the-shelf lossy compressors distort such essential features.

Even the tiniest variations, however small the compression error bound is, could induce significant alternations in the MS segmentation, thereby impacting scientists' understanding of the data. 
In Figure~\ref{fig:intro}, we compress a molecular dynamics dataset using SZ3 and ZFP with relative error bounds ranging from $10^{-5}$ to $10^{-2}$, which all introduced discrepancies of topological segmentations induced by MS complexes up to 100\%; the specific metric measures the discrepancies of segmentations by the percentage of points with a wrong segmentation ID, as explained later in this paper.

This research focuses on preserving Morse-Smale (MS) segmentations~\cite{MSS} in lossy compression workflows; such segmentations provide a preview of the Morse-Smale complex.  MS segmentation consists of two significant components of the full MS complex: the extrema designated for each MS complex region and the boundaries separating adjacent MS segmentations. Due to the high computational complexity of the MS complex, MS segmentations are used in diverse applications because they offer a cost-effective way to visualize topological segmentation for visualization and analyses.

To tackle inconsistencies of topological segmentations, we introduce a novel edit-based paradigm for preserving MS segmentations in error-bounded lossy decompressed data in 2D/3D piecewise linear (PL) scalar fields.
Unlike existing efforts that preserve other topological features by tailoring a specific compressor~\cite{yan2023toposz, LiangDCRLOCPG23, 9086223, Liang24}, our method directly edits compression outputs, making it theoretically applicable to any error-bounded lossy compressors that applications already trust and engage. These edits also allow us to adjust and verify outputs recurrently without repeatedly calling compressors with different input parameters.
Specifically, we identify a subset of data that requires edits, such that the MS segmentation of the decompressed data remains identical to that of the original dataset, while also guaranteeing the global error bound. Since each MS segmentation region consists of a pair of extrema and the points included in the region, our method mainly focuses on preserving these two features by alternatively fixing extrema and integral lines until convergence within finite iterations of the loop: (1) computing the MS segmentation of the current data, (2) identifying the false data points (false critical or regular points), and (3) fixing the false data points until one can reconstruct the exact MS segmentation with the edits during decompression. 

We address the performance and scalability of our method with shared-memory/GPU parallelism. While the efficient parallel computation of MS segmentation/complex is already a known challenge~\cite{MSS}, the full realization of our vision requires careful parallelization design. Specifically, a key challenge is handling read-write and write-write conflicts when multiple threads attempt to modify the data concurrently. For example, if a vertex is adjacent to two critical points, during the iteration to fix multiple critical points, two parallel threads could simultaneously change the vertex's value; we use atomic intrinsics to avoid such conflicts while maintaining high scalability. 
As part of our implementation, we also introduce a GPU version of Maack et al.~\cite{MSS} to boost the overall performance of our workflow.
We perform a comprehensive performance analysis with CPUs and GPUs on a compute node of Lawrence Berkeley's Perlmutter supercomputer. 
In summary, the contributions of this paper are multifold:
\begin{itemize}
    \setlength{\parskip}{0pt}
    \item A novel edit-based paradigm for preserving MS segmentation within error-bounded lossy decompressed data in 2D/3D piecewise linear scalar fields, theoretically applicable to any existing error-bounded lossy compressors;
    \item Efficient shared-memory/GPU parallelism that significantly accelerates individual components of our algorithm while addressing read-write/write-write conflicts; 
    \item Comprehensive evaluation of our method across various datasets, two off-the-shelf base compressors (SZ3 and ZFP), and parallel performance on both CPUs and GPUs. 
\end{itemize}

\section{Related Work}
We review related work on error-bounded lossy compression, topology-preserving compression, and Morse-Smale complexes.

\subsection{Error-Bounded Lossy Compression}
Error-bounded lossy compressors can significantly reduce data by allowing bounded pointwise error between the decompressed and original data within a user-specified threshold.  Note that the number of points/vertices remains constant from the original to the compressed data.
However, few existing lossy compressors consider topology features in the decompressed data, which will impact disciplines that require post hoc analysis of the topological structure, thereby introducing deviations in the analytical results.

Error-bounded lossy compressors can be divided into prediction- and transformation-based methods. Examples of prediction-based compressors include the SZ series~\cite{sz, sz3, Tao_2017, Kai2021, Kai2020, Di2018,Di2016}.  For example, SZ1.4~\cite{Tao_2017} uses the Lorenzo predictor combined with linear-scaling quantization to convert prediction residuals into integers, which are then encoded with customized Huffman coding and lossless compressors like ZSTD~\cite{zstd} and GZIP~\cite{GZIP}. 
QoZ~\cite{Qoz} is an optimization of SZ3, focusing on improving the quality of decompressed data under dynamic metrics with parameter auto-tuning. It automatically adjusts the compression based on user-specified quality objectives. 
FPZIP~\cite{FPZIP} allows a specified number of bit planes to be ignored, making the data distortion controllable on demand. 
AE-SZ~\cite{liu2023exploring} and SRNN-SZ~\cite{liu2023srnsz} are examples of prediction-based compressors that use neural networks.

Transformation-based lossy compressors first transform data into an alternative representation, such as wavelet transformation and tensor decomposition, then compress data in the transformed domain. 
For example, ZFP~\cite{zfp} uses a custom orthogonal block transform to decorrelate data within blocks, transforming original data into sparsely distributed coefficients. 
These coefficients are then encoded for efficient compression, leveraging the reduced complexity of the transformed data to enhance compression efficiency. 
TTHRESH~\cite{TTHRESH} is a transform-based lossy compressor that uses bit-plane, run-length, and arithmetic coding to compress the transform coefficients of the higher-order singular value decomposition.
SPERR~\cite{SPERR} is another transform-based compressor that uses the CDF9/7 discrete wavelet transform. 

Metrics commonly used to evaluate the quality of lossy compression include mean square root error (RMSE), peak noise-to-signal ratio (PSNR), and structural similarity (SSIM); see Di et al.~\cite{di2024survey} for a comprehensive review of this topic.  We further describe metrics for evaluating MS segmentations later in this paper.

\subsection{Topology Preservation in Lossy Compression}

Topology preservation is an emerging topic in the context of error-bounded lossy compression; to our knowledge, our work is the first attempt toward preserving MS complexes by maintaining the consistency of topological segmentations.
For scalar fields, previous studies primarily focus on contour/merge tree preservation. Yan et al.~\cite{yan2023toposz} introduced TopoSZ to enhance the SZ 1.4 compression algorithm by integrating topological constraints informed by segmentations induced by contour trees.  
Soler et al.~\cite{soler2018topologically} developed a topology-controlled compression scheme to adaptively quantize data in individual topological features to preserve the persistence diagram subject to a persistence simplification threshold.  
For vector fields, researchers have attempted to retain critical points, yet the preservation of topological segmentations is studied empirically.  For example, Liang et al.~\cite{9086223, LiangDCRLOCPG23} presented a methodology for preserving critical points in piecewise linear and bilinear vector fields. 

\subsection{Morse-Smale Complexes and Segmentations}

We summarize related work in MS complexes and segmentations and leave a detailed review of key definitions in the next section. 
MS complexes are a well-studied topological descriptor researched by the topological data analysis (TDA) and visualization communities.  Constituents of MS complexes include critical points (maxima, minima, and saddles) and separatrices that connect saddles and extrema. The separatrices also partition the input manifold into regions with monotonous subregions, often referred to as MS segmentations.

In general, two flavors exist for MS complex computation: piecewise linear (PL) Morse theory~\cite{cp,msc2} and discrete Morse theory~\cite{FORMAN199890}.  We refer readers to Lewiner et al.~\cite{LEWINER2013609} for a comprehensive review and comparison between the two approaches.  Our work primarily relies on the PL-based MS segmentation computation, and we formally review key assumptions and concepts of PL Morse theory in the next section.  

With the PL Morse theory, one can further divide algorithms into boundary- and region-growing-based approaches.  For boundary-based algorithms, Edelsbrunner et al.~\cite{msc2} first introduced the MS complex for PL 2-manifolds and 3-manifolds~\cite{plmsc3d}.  Gyulassy et al.~\cite{Gyulassy} introduced the region-growing algorithm for analyzing MS complexes, which is scalable for 3D or higher dimensions. Banchoff et al.~\cite{cp} explored the critical points on PL 2-manifolds by analyzing the paths of the steepest ascent and descent.  

With Forman's discrete Morse theory, Fugacci et al.~\cite{fugacci2018computing} proposed an efficient algorithm for large and high-dimensional simplicial complexes. Gyulassy et al.~\cite{Gyulassy1} introduced an algorithm that computes MS complexes across different data scales and dimensions, which was later extended to a parallel computing framework~\cite{Gyulassy2}.

In terms of parallel computation of MS complexes and segmentations, Beucher et al.~\cite{Beucher} utilized watershed transformations to analyze and construct MS complexes in the context of image processing and data analysis, which was optimized by Gabrielyan et al.~\cite{Gabrielyan} by leveraging GPUs. 
Yeghiazaryan et al.~\cite{Yeghiazaryan} combined path simplification with watershed transformations for efficiency.

\section{Background: Morse-Smale Segmentation in Piecewise Linear Scalar Fields} 
\label{sec:background}

We review key concepts in piecewise linear MS segmentations (PLMSS)~\cite{MSS}. Formally, for a piecewise linear function (represented with a triangular/tetrahedral mesh) with a distinct value $f_i$ for each vertex $i$, PLMSS assigns a pair of minimum and maximum labels $\langle m_i, M_i \rangle$ by tracing integral lines along the steepest ascending/descending directions along mesh edges.

\begin{wrapfigure}{R}{1.05in}
    \vspace{-0.2in}
    \includegraphics[width=0.9\linewidth]{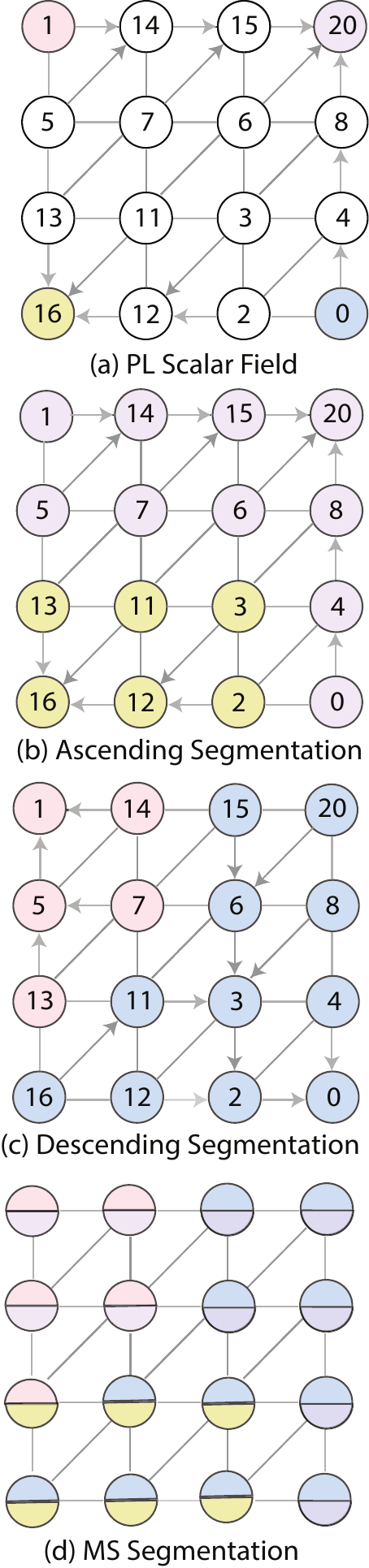}
    \caption{Constituents of MS segmentations.}
    \label{fig:mss}
\end{wrapfigure}

\textbf{Critical points} in PL scalar fields reside on the vertices of the underlying triangulation, and they are extracted and classified based on the function values of neighboring vertices.
The computation of PLMSS concerns only minimum and maximum.  For example, in Figure~\ref{fig:mss}(a), vertices with scalar values 0 and 1 are minimum because their values are the lowest among their neighbors; likewise, vertices with 16 and 20 are maximum.

\textbf{Integral lines}.
In PL scalar fields, integral lines of gradient vector fields are constructed by monotonic paths consisting of edges in the triangulation.  For example, in Figure~\ref{fig:mss}(b), $5\rightarrow 14\rightarrow 15\rightarrow 20$ is an integral line that extends toward the highest/lowest adjacent vertices until a maximum/minimum is met.  We further distinguish backward and forward integral lines based on the ascending/descending direction an integral line is tracing toward.

\textbf{Ascending/descending segmentations}. Once integral curves are traced, one can find the minimum/maximum that a vertex $i$ flows to and further define MS segmentation in PL scalar fields.  Starting from $i$, we denote its converging critical points as the maximum label $M_i$ and minimum label $m_i$ following the ascending and descending integral lines.   
For example, in Figure~\ref{fig:mss}, nodes in b and c, respectively, are colored by maximum and minimum labels; nodes in d are colored by both maximum and minimum labels.

Compared with MS complexes, PLMSS is suitable for applications that do not need to compute full MS complexes.  
The most notable difference is that PLMSS does not concern saddle points, leading to two limitations as reviewed by Maack et al.~\cite{MSS}: (1) PLMSS cannot capture saddle-saddle separatrices, (2) downstream tasks (e.g., persistence simplification) that rely on full MS complexes do not apply to PLMSS. That said, PLMSS offers a fast and practical tool to understand scalar fields.

\section{Problem Statement}

We formulate the preservation of MS segmentations in 2D and 3D piecewise linear scalar fields.  The inputs of our algorithm include both original data $f$ and decompressed data $\hat{f}$ with exactly the same number of vertices, assuming both data versions are available at the compression time.  We assume all scalar field data are Morse; that is, for any two vertices $i$ and $j$, we have $f_i\neq f_j$; otherwise, we use simulation of simplicity (SoS)~\cite{Edels94} to handle non-Morse regions for real-world data.
The outputs of our algorithm are a series of edits $\{\delta_i\}$; with the edits, one can derive the final edited value at vertex $i$ as $g_i=\hat{f}_i+\delta_i$.  
The edited value shall satisfy the following constraints: 

\textbf{Preservation of the global error bound}.  We must guarantee that the final edited value is subject to the user-prescribed absolute error bound $\xi$, that is, $\left|f_i-g_i\right|\leq\xi$.

\textbf{Preservation of MS segmentations}.  Let $M$ and $m$, respectively, denote the maximum and minimum labels in MS segmentations of the original data, and let $\hat{M}$ and $\hat{m}$ denote the counterparts in the decompressed data.  This research aims to precisely align $\hat{M}$ with $M$ and $\hat{m}$ with $m$.  That is, for any vertex $v_i$, we have $M_i=\hat{M}_i$ and $m_i=\hat{m}_i$.

\begin{figure*}[ht]
     \centering 
     \includegraphics[width=\textwidth]{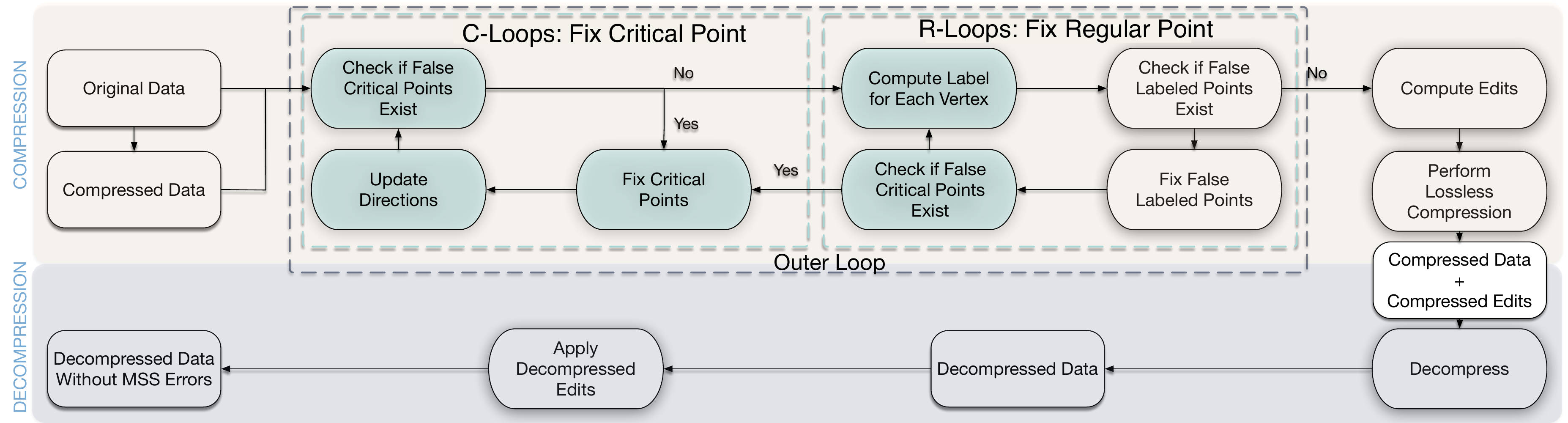}
     \caption{Compression (top) and decompression (bottom) workflows.  Our algorithm derives vertexwise edits with
     two distinct loops: (1) C-loops that iteratively fix all false critical points and (2) R-loops that fix all false labeled regular points.  
     C- and R-loops execute alternatively until there is no false critical/regular point. 
     The edits, which are losslessly stored, are used to correct MS segmentations in the decompression stage.
     }
     \label{fig:workflow}
 \end{figure*}      

Preserving PLMSS implies that all extrema are preserved without any false positives or negatives.  Under the premise that the location and type of critical points in the decompressed data are identical to those in the original data, to ensure that $\hat{M}$ is precisely equal to $M$, it is necessary to align the label of each regular point $i$ in the decompressed data with its corresponding label in the original data. 

\section{Methodology} \label{sec:method}
This section describes the theoretical workflow of computing edits for preserving MS segmentations, as illustrated in Figure~\ref{fig:workflow}.  The derivation of vertexwise edits is an iterative process involving two distinct loops: the critical point loops (C-loops, Section~\ref{sec:fixcp}) and the regular point loops (R-loops, Section~\ref{sec:fixreg}).  The rationale for alternating the two loops is that (1) the correctness of critical points is necessary for fixing regular points, and (2) fixing regular points may introduce new false critical points to be fixed further which are discussed in Section~\ref{sec:convergence}.

\subsection{Critical Point Loops (C-Loops)}\label{sec:fixcp}

We fix four types of false critical points: 
false positive maxima (FPmax), false positive minima (FPmin), false negative maxima (FNmax), and false negative minima (FNmin). A sublevel loop handles each false type; for example, the FPmax loop executes multiple times until no FPmax exists, followed by the FPmin loop. A C-loop sequentially executes the four subloops in each iteration and exits when no false critical point remains. 

Readers may skip the mathematical reasoning below, but the key to (sub)loop convergence is incurring changes that only decrease (or keep) scalar values at each iteration.  That is, denoting $g_i^{(k)}$ as the edited value at vertex $i$ during the $k$th iteration ($k\in\mathbb{I}$), we have
\begin{equation}\label{eq:iter}
     \hat{f}_i = g_i^{(0)} \geq \cdots \geq g_i^{(k)} \geq g_i^{(k+1)} \geq \cdots > f_i-\xi, 
\end{equation}
where $\hat{f}_i$ is the initial decompressed value,
and $g_i^{(k)}$ progressively approaches to the lower bound $f_i-\xi$.  As such, one can make the relationship between scalars on neighboring vertices consistent with that of the original data; formally, for arbitrary two vertices $i$ and $j$, we have 
\begin{lemma}\label{lemma1}
One can find a finite number $k$ of iterations such that $g_i^{(k)} < g_j^{(k)}$, if initially $g_i^{(0)} > g_j^{(0)}$ and $f_i < f_j$.
\end{lemma}

\noindent This holds because $f_i-\xi = \lim\limits_{k\to\infty}g_i^{(k)} < \lim\limits_{k\to\infty}g_j^{(k)} = f_j-\xi$. Backed by Lemma~\ref{lemma1}, we design decreasing edits to fix four false cases.  Each decreasing edit applies to the vertex with a false critical point (FPmax or FNmin) or a neighbor vertex of the false critical point (FPmin or FNmax), as exemplified below.

\subsubsection{False Positive Maximum}\label{sec:fpmax}

\begin{definition}[FPmax] A maximum $g_i$ is false positive if $g_i>g_j$ for all neighboring vertices of $i$, 
but one can find at least one neighbor $j$ such that $f_i<f_j$.\label{def1}
\end{definition}

\noindent We construct a sequence of iterations for all vertices, following Equation~\eqref{eq:iter}, to iteratively eliminate all FPmax: 
\begin{equation}\label{eq:fpmax}
    g_i^{(k+1)}:= \left\{ 
    \begin{array}{ll}
        \nicefrac{\left(g_i^{(k)} + f_i - \xi\right)}{2}, & \mbox{~if $g_i^{(k)}$ is FPmax} \\
        g_i^{(k)}, & \mbox{~otherwise.}
    \end{array}
    \right.
\end{equation}
Note that $g_i^{(k)}$ monotonically decreases 
toward the lower bound $f_i-\xi$ as $k$ increases.  
We speculate three possible outcomes:
\begin{itemize}
    \setlength{\parskip}{0pt}
    \item Case I: $g_i^{(k)}$ remains an FPmax, requiring at least another iteration 
    \item Case II: $g_i^{(k)}$ becomes non-maximum without introducing any new FPmax; for the moment, $g_i^{(k)}$ is fixed; 
    \item Case III: $g_i^{(k)}$ becomes non-maximum but introduces at least one new FPmax $g_j^{(k+1)}$ at a neighboring vertex $j$; 
\end{itemize}
Cases I and II are trivial. 
In Case III, the newly introduced FPmax at $j$ will be eventually fixed in later iterations without making $i$ an FPmax again.  Specifically, per Definition~\ref{def1}, one can find a neighboring vertex $j$ 
that is ascending in the original data ($f_j<f_{j^\prime}$); no matter if $j^\prime$ is $i$ or not, with additional iterations with whichever cases, per Lemma~\ref{lemma1}, neither $i$ nor $j$ will become FPmax with additional finite iterations.

\begin{figure}[!hbt]
    \centering
    \includegraphics[width=\linewidth]{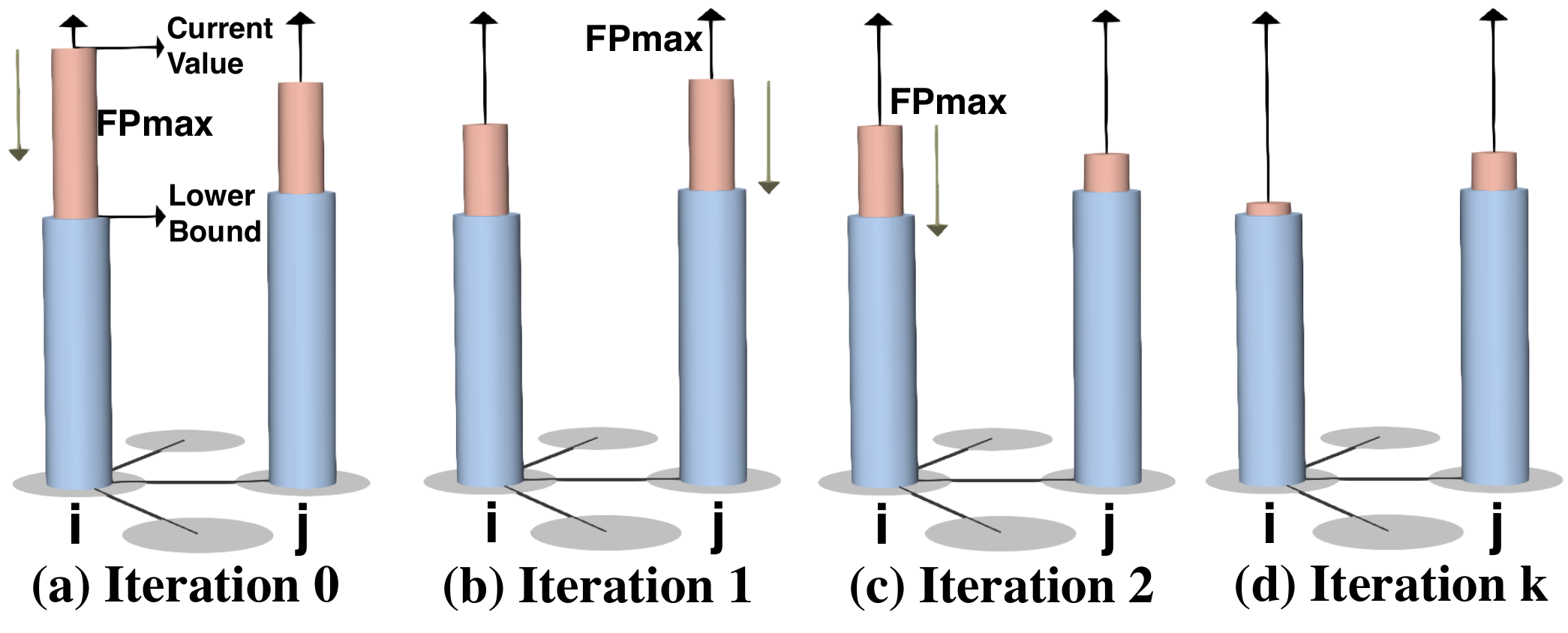}
    \caption{Fixing an FPmax at vertex $i$.  The height of the blue cylinder above each vertex represents its lower bound ($f - \xi$), and the height of the pink cylinder represents its current value.}
\label{fig:explai}
\end{figure}

We use Figure~\ref{fig:explai} to help further understand Case III.  Initially, vertex $i$ is an FPmax, and the decompressed value is greater than all its neighbors (shown in a).  However, in the original data, we have at least one neighbor $j$ such that $f_j>f_i$.
To turn $i$ back into a regular point, in the first iteration, we decrease the $g_i$ 
such that $g_i^{(1)} < g_j^{(1)}$.
However, $j$ could become a new FPmax due to the decrement of $g_i$ (shown in b).  In the next iteration (c), as we fix the new FPmax $j$ by decreasing $g_j$, 
vertex $i$ could turn into FPmax again. The false positives may recur up to a finite number of iterations but will guarantee to vanish because we incrementally drive $g_i$ and $g_j$ closer to their lower bounds. Note that $f_i < f_j$ and the lower bound $f_i - \xi < f_j - \xi$, with a finite number $k$ of iterations, we have $g_i^{(k)} < f_j - \xi < g_j^{(k)}$ (shown in d).  At this time, vertex $i$ will no longer become a new false positive maximum.

\subsubsection{False Positive Minimum}

\begin{definition}[FPmin] A minimum $g_i$ is false positive if $g_i<g_j$ for all neighbors $j$ but one can find at least one neighbor $j$ such that $f_i>f_j$.\label{def2}
\end{definition}

\noindent Unlike fixing FPmax, assuming $g_i^{(k)}$ is FPmin, we decrease the value at the \emph{ascending neighbor} $j$, such that $g_j^{(k)}$ is the maximal among the neighbors of $i$, and we have 
\begin{equation}\label{eq:fpmin}
    g_j^{(k+1)}:= \left\{ 
    \begin{array}{ll}
        \nicefrac{\left(g_j^{(k)} + f_j - \xi\right)}{2}, & \mbox{~if $g_j^{(k)}$ is maximal among $i$'s neighbors} \\ 
        g_j^{(k)}, & \mbox{~otherwise.}
    \end{array}
    \right.
\end{equation}
While one could alternatively fix the FPmin by increasing the value on the same vertex, we impose decreasing edits across all types of false critical points to guarantee convergence across the iterative workflow; otherwise, incompatible strategies may not lead to convergence.  
Later in this section, we demonstrate how an FPmin is fixed amid a complex process and discuss the convergence.

\subsubsection{False Negative Maximum/Minimum}

\begin{definition}[FNmax/FNmin] A non-maximum (or non-minimum) $g_i$ is false negative if $f_i>f_j$ (or $f_i<f_j$) for all $j$ in neighbors of $i$. 
\end{definition}

\noindent Specifically, for an FNmax $g_i$, we reduce its ascending neighbor's value: 
\begin{equation}\label{eq:fnmax}
    g_j^{(k+1)}:= \left\{ 
    \begin{array}{ll}
        \nicefrac{\left(g_j^{(k)} + f_j - \xi\right)}{2}, & \mbox{~if $g_j^{(k)}$ is maximal among $i$'s neighbors} \\ 
        g_j^{(k)}, & \mbox{~otherwise.}
    \end{array}
    \right.
\end{equation}
For an FNmin $g_i$, we decrease its own scalar value:
\begin{equation}\label{eq:fnmin}
    g_i^{(k+1)}:= \left\{ 
    \begin{array}{ll}
        \nicefrac{\left(g_i^{(k)} + f_i - \xi\right)}{2}, & \mbox{~if $g_i^{(k)}$ is FNmin} \\
        g_i^{(k)}, & \mbox{~otherwise.}
    \end{array}
    \right.
\end{equation}
Note that both strategies comply with Equation~\eqref{eq:iter} so that the iterations are provably convergent with finite iterations.

\subsection{Regular Point Loops (R-Loops)}\label{sec:fixreg}

Once all false critical points are fixed, the next step is to fix all regular points' maximum and minimum labels in the decompressed data.
For a falsely labeled regular point, our method involves three steps in each iteration: (1) compute the ascending/descending integral lines in the current edited data $g^{(k)}$, (2) locate the \emph{troublemaker}, which is defined as as the first occurrence of discrepancy along the integral lines (as illustrated in Figure~\ref{fig:fixpath}(b)), 
(3) reroute the troublemaker by an edit.  Specifically, for a falsely labeled regular point $i$ with a wrong minimum (maximum), let $v_t$ represent the troublemaker's descending (ascending) neighbor,
 we decrease the value at $v_t$:
\begin{equation}
g_{v_t}^{(k+1)} := \nicefrac{\left(g_{v_t}^{(k)}+f_{v_t}-\xi\right)}{2}.
\label{eq:bias5}
\end{equation}
Note that edits in an R-loop may introduce new false critical points; in this case, we must return to the C-loop to address the new false cases.  Similar to C-loops, edits in R-loops always decrease decompressed values such that no new false critical points are introduced after a finite number of iterations.

\subsection{Convergence of Alternating C- and R-loops}\label{sec:convergence}

We alternatively execute C- and R-loops until all false critical/regular points vanish.  The workflow converges because all edits progressively decrease decompressed values towards the lower bound.  The convergence is backed by Lemma~\ref{lemma1}; because the inconsistent ordering of scalar values may cause changes to ascending/descending directions and lead to false cases, the inconsistencies will guarantee to vanish with a finite number of iterations with our decreasing edit strategies.

\begin{figure}[!hbt]
    \centering
     \includegraphics[width=\linewidth]{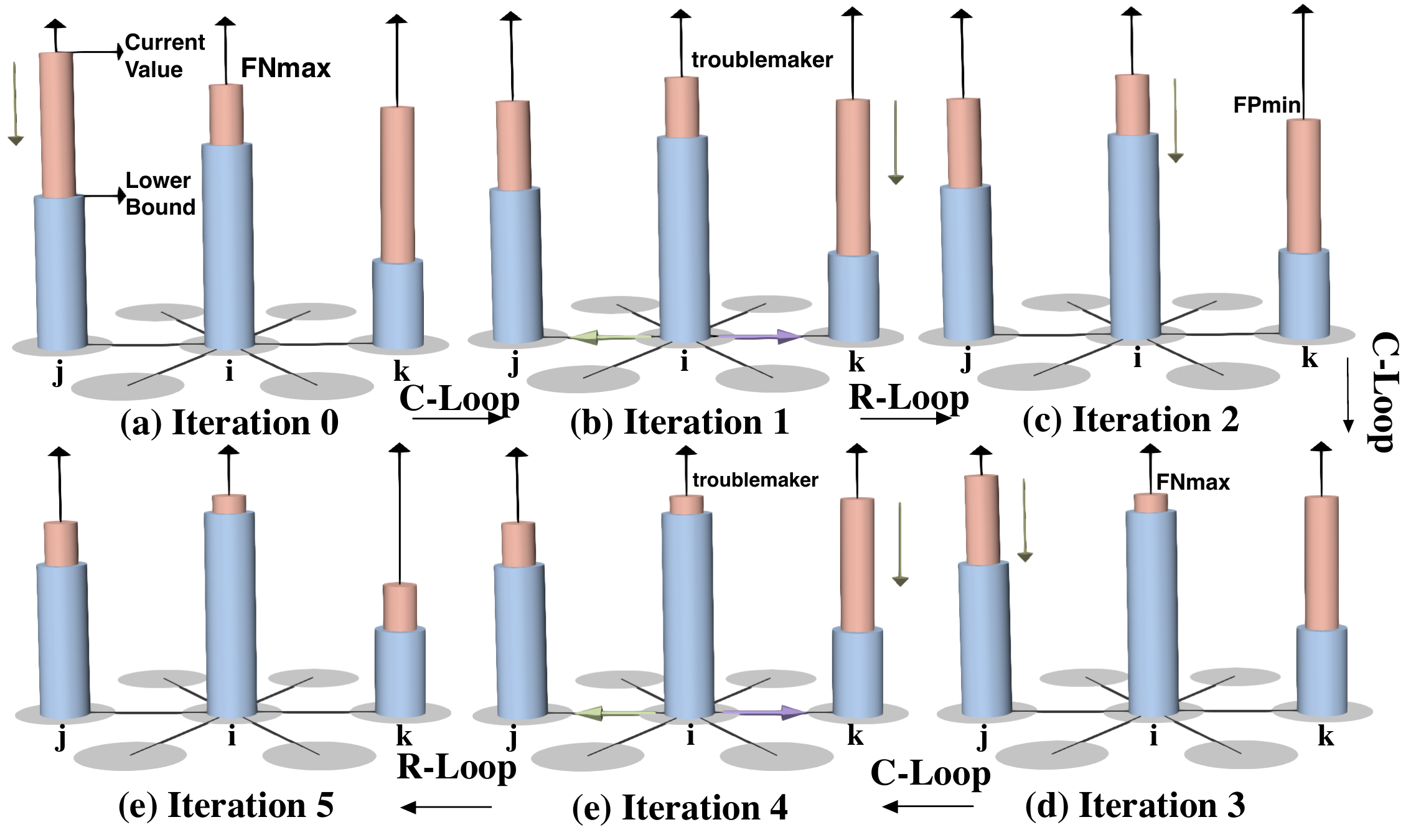}
    \caption{Fixing an FNmax (and subsequently, a troublemaker, FPmin, FNmax, and another troublemaker) across multiple C- and R-loops. The purple and green arrows in (b) and (d) show the ascending neighbor of vertex $i$ in the original and decompressed data, respectively.}
\label{fig:fixpath}
\end{figure}

Figure~\ref{fig:fixpath} demonstrates a specific case of alternating C- and R-loops to fix an FNmax.  We have $f_i>f_j>f_k$ in the original data, but initially $g_j>g_i>g_k$.  The FNmax at vertex $i$ is fixed in the first C-loop by decreasing the neighbor $j$.  However, $i$ becomes a troublemaker that causes false regular points.  After the troublemaker is fixed by decreasing its neighbor $k$, $k$ becomes an FPmin, thus triggering another C-loop.  The second C-loop fixed the FPmin by decreasing $i$, which caused a new FNmax case at $i$.  With two additional C- and R-loops for newly introduced false cases, we have no more false cases to address.

\section{Parallel Computation and Compression of Edits} \label{gpu}
This section describes the shared-memory/GPU parallelism that significantly accelerates individual components of our algorithm.

\subsection{Parallel C-Loop}\label{cloop}

We parallelize all subroutines in C-loops, including (1) finding false critical points, (2) fixing false critical points, and (3) updating directions for each vertex, as highlighted in Figure~\ref{fig:workflow}. 

\textbf{Finding false critical points} 
checks each vertex in parallel to see if it is a false critical point.  
We assign each thread with a vertex; if the vertex is false critical, we push the case into a (lock-free) stack for further processing.  
For both CPU and GPU architectures, we implement the lock-free stack with \texttt{atomicAdd} operations with a pre-allocated buffer. For each push, we record the current stack height as $h$ while increasing the height by one; this fetch-and-add operation guarantees to avoid race conditions. We then write the false case (represented by vertex ID and false type) into the $h$th position of the buffer.

\textbf{Fixing false critical points} derives and applies edits for each false critical point with a thread.
Because an edit may change the scalar value of the false critical point or one of its neighbors, multiple threads may edit the same value and cause write-write conflicts. In a conflict, only one edit will be applied when multiple threads preemptively edit the same vertex, but the specific choice made by hardware is random; we call it \emph{preemptive mode}. The preemptive mode would still converge because the missing edits would be applied by a later iteration, albeit of random execution orders. To make executions consistent, we incorporate atomic compare-and-swap (\texttt{atomicCAS}) operations in our CPU and GPU implementations.  Specifically, the operation compares the incoming edit with the current edit value and arbitrarily applies the most significant edit, making execution orderings deterministic.

\textbf{Updating directions}.  We assign each vertex to a thread to update its ascending/descending neighbor for identifying false cases.  This step involves a local comparison of scalar values between neighbors.

\subsection{Parallel R-Loop}

The only component we currently parallelize in the R-loop is the computation of MS segmentation, which dominates the execution time. 
We implemented path compression~\cite{pathcompression} used by Maack et al.~\cite{MSS}, for the computation of Morse-Smale segmentation.
On GPUs, taking ascending segmentation as an example, path compression involves utilizing a (lock-free) list to track regular points that have not yet found their maximum. For these points, the method seeks its largest neighbor, updating each point's value in the list to that of its largest neighbor's largest neighbor. If the value of a point $v$ remains unchanged after the update, meaning the largest neighbor's largest neighbor is itself, it indicates that $v$ has been assigned to a maximum. The iteration concludes once every point has successfully determined its maximum.

\subsection{Lossless Compression of Edits}\label{sec:edits}

Once all iterations converge and exit with no false critical/regular points, the last step is to store the edits compactly as metadata appended to the lossy compressor's outputs. Each edit $\delta_i$ is represented with a key-value pair, the key being the vertex index and the value being the floating-point representation of the edits (See Supplementary Material for a visualization of the spatial distribution of edits).  
We compress the indices and edit values separately.  Regarding the indices, we first sort them in ascending order and compress the differential sequence,
because storing the differentials makes it possible to maximize the use of run-length encoding (RLE) and Huffman coding before offloading the edits to a lossless compressor such as ZSTD~\cite{zstd} or GZIP~\cite{GZIP}.

\section{Evaluation of Edited Data}
We evaluate our method with datasets from diverse applications, ranging from climate and cosmology to combustion, by measuring the accuracy in MS segmentations based on two state-of-the-art error-bounded lossy compressors, SZ3 and ZFP.  Evaluation metrics are as follows; detailed descriptions of datasets and metrics are in Supplementary Material.
\textbf{Overall Compression Ratio (OCR)} is the compression ratio after the combination of edits, calculated by the combined size of compressed edits and data over the original data size. A higher OCR indicates more effective compression, resulting in a smaller compressed file.
\textbf{Edit ratio} quantifies the proportion of data points that are edited to fully preserve the MSS in the decompressed data. It is calculated as the number of modified data points divided by the total number of data points in the decompressed data.
\textbf{Overall bit rate (OBR)} represents the average number of bits required to encode each data point after compression (after the combination of compressed data and edits). It is calculated by dividing the total number of bits used by the total number of data points. Lower bit rates indicate more efficient compression.
\textbf{Right labeled ratio} is the percentage of points in the data with correct MSS labels, calculated by the number of right labeled points over the number of points in the data.
\textbf{PSNR distortion} refers to the trade-off between the bitrate and PSNR. It describes how the bitrate affects the PSNR of the decompressed data, where a higher bitrate generally results in a higher PSNR, indicating better quality. This is typically represented as a curve with the x-axis showing the bitrate and the y-axis showing the PSNR of the decompressed data.
\textbf{MSS distortion} is similar to PSNR distortion, with the y-axis changed to the right labeled ratio. It reflects the trade-off between the bitrate and the degree of preservation of the MSS. 

\subsection{Features of Interest Preservation}

We exemplify features of interest characterized by MS segmentations and how off-the-shelf lossy compressors distort the scientific insights. Examples below are based on SZ3 with a 1\% error bound. 

\begin{figure}[hbt!]
\centering
    \includegraphics[width=\linewidth]{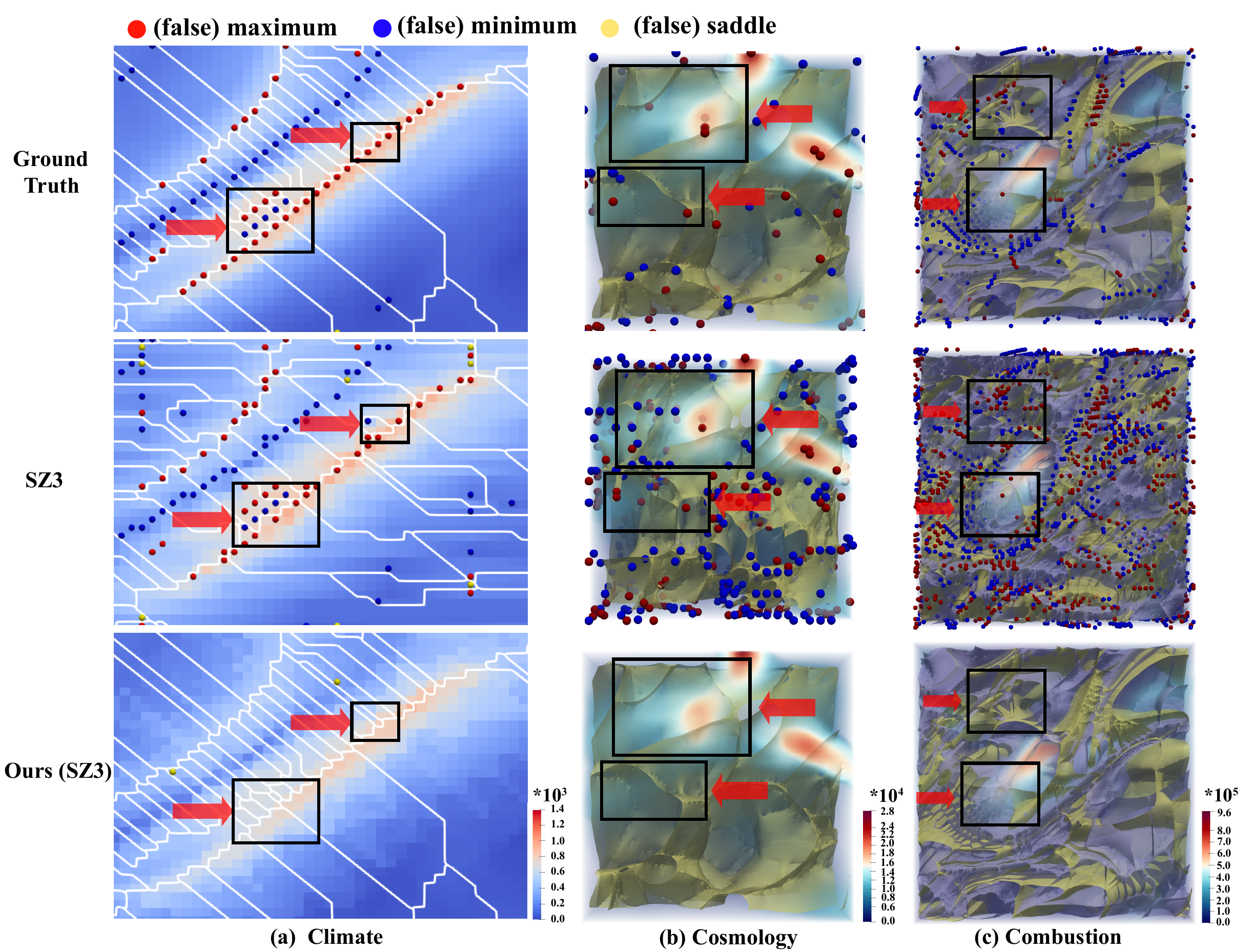}
    \caption{Feature preservation in SZ3-compressed data for (a) climate, (b) cosmology, and (c) combustion.}
\label{fig:case_study} 
\end{figure}

\textbf{Climate.}  Distortions in \textbf{Atmospheric Rivers} (ARs), based on feedback from domain scientists, could significantly impact scientists' understanding of AR formation and development, potentially leading to inaccurate evaluations of ARs' impacts on precipitation and flooding in North America.  
Scientists characterize AR skeletons by the boundaries of descending segmentations of the Integrated Vapor Transport (IVT) field~\cite{Lan_ivt}.  As shown in Figure~\ref{fig:case_study}(a), SZ3 led to distorted segmentation boundaries of the AR and surrounding regions. 

\textbf{Cosmology}. Scientists characterize \textbf{cosmic walls} in dark matter distributions by ascending segmentation boundaries~\cite{Sousbie_2011}.  Distorted sheet structures of walls, as exemplified in a zoomed-in view of the Nyx data (a $50^3$ crop of the entire volume) compressed by SZ3, could lead to incorrect separation between voids and potentially impact the understanding of the cosmic web as shown in Figure~\ref{fig:case_study}(b).

\textbf{Combustion}. Scientists use MS segmentations to identify \textbf{flame structures} to understand the reaction, mixing, and quenching in the burning dynamics~\cite{Bremer_combustion, Bremer_flames}.  We demonstrate feature distortion in a zoomed-in region ($100^3$) in Figure~\ref{fig:case_study}(c). SZ3 introduced false features and omitted essential separatrices, potentially leading to incorrect identification of burning zones and understanding of fuel consumption areas.

\begin{figure*}[htb!]
     \centering 
     \includegraphics[width=\linewidth]{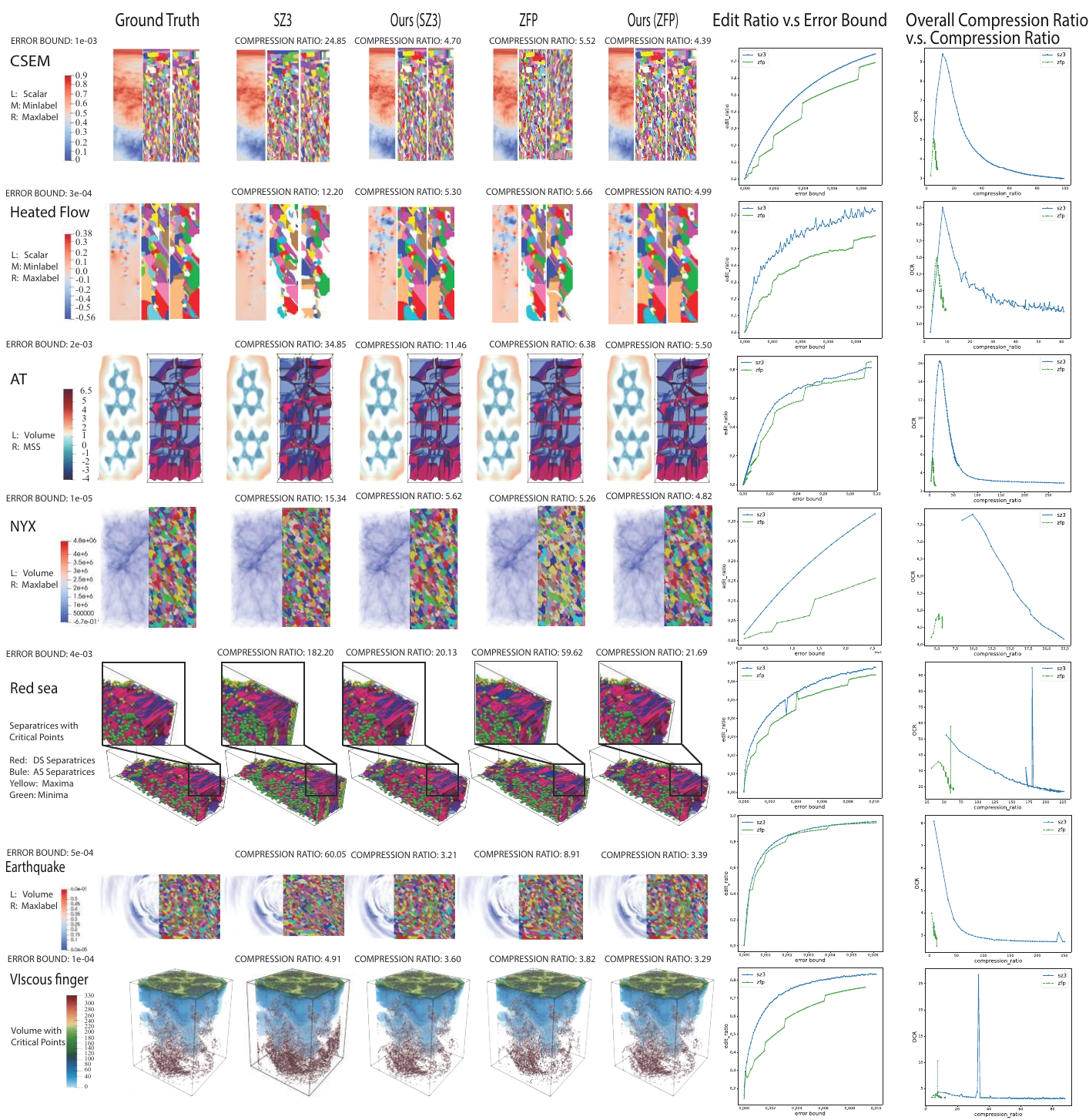}
     \caption{Fixed-error-bound comparison of lossy compressors on different datasets.}
     \label{fig:fixed_error/}
\end{figure*}
\subsection{Fixed-Error-Bound Comparison} \label{error}

We demonstrate that our method can fix MS segmentations from arbitrary outputs with different compressors and error bounds, as shown in Figure~\ref{fig:fixed_error/}.  
We observe that the size of the edits required is roughly proportional to the magnitude of the global error bound.  This is reasonable because as the global error bound increases, implying that more errors are introduced into the data, the error rate in the MSS escalates, thereby necessitating a greater number of edits. For more complex datasets, such as CSEM, Nyx, and viscous fingering, the edit ratio is noticeably higher than other datasets.  Another observation is that edits for ZFP are generally lower than those for SZ3, while OCR is higher overall than ZFP because ZFP's original compression ratio is lower than SZ3.

\begin{figure*}[htb!]
     \centering 
     \includegraphics[width=\linewidth]{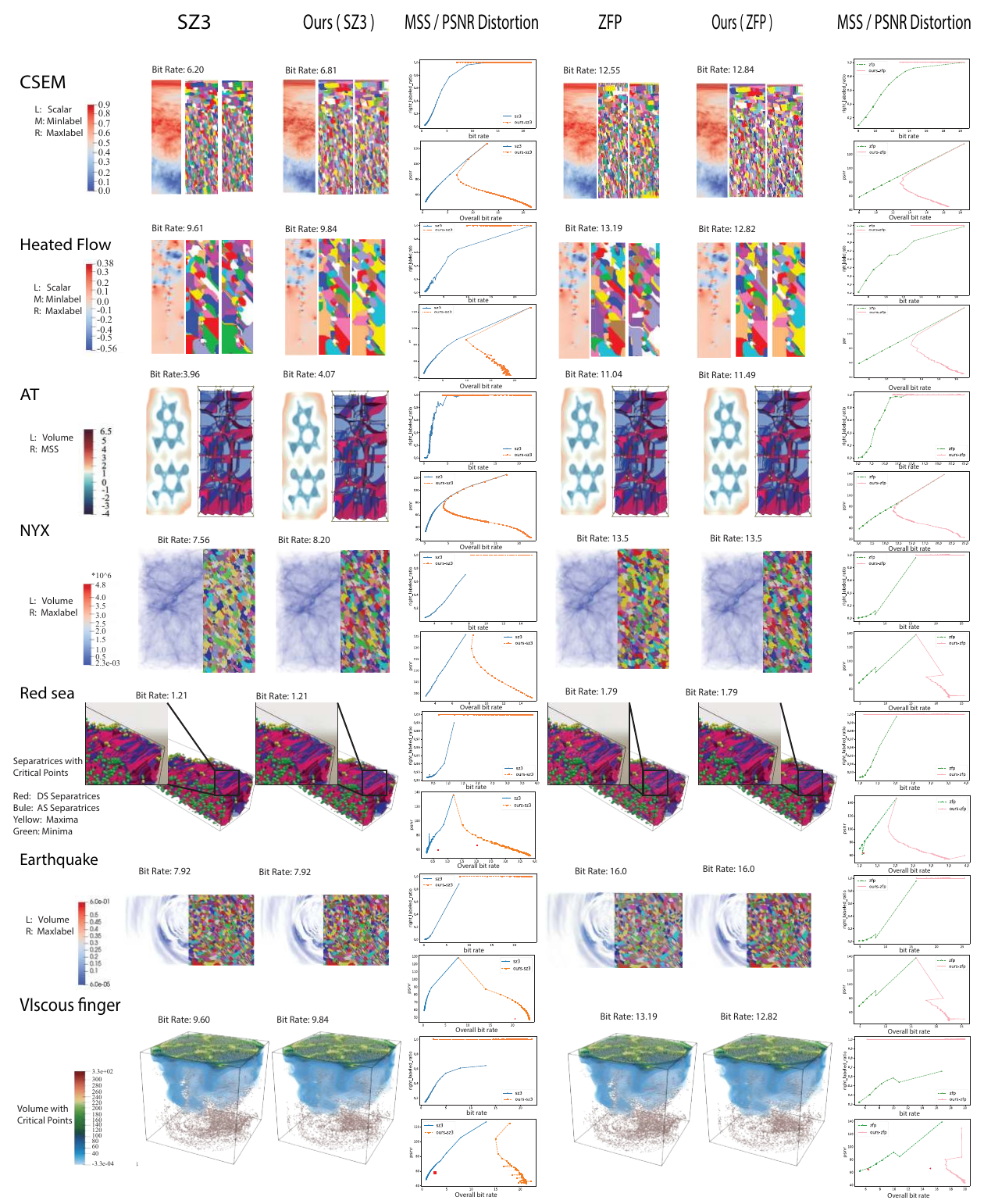}
     \caption{Fixed-bit-rate comparison of lossy compressors on fixed bit rate on different datasets.} 
     \label{fig:fixed_bitrate/}
\end{figure*}
\subsection{Fixed-Bit-Rate Comparison}

We evaluate our method across various datasets to identify the optimal overall bit rate (after the combination of edits) achievable on SZ3 and ZFP, as shown in Figure~\ref{fig:fixed_bitrate/}. 
We run an ensemble of experiments with various error bounds and find relatively the same (overall) bit rate with both SZ3 and ZFP to evaluate the preservation of MSS. 
With comparable bitrates, the original decompressed data from SZ3/ZFP exhibit varying degrees of MSS distortion across different datasets.  Despite a certain bitrate increase due to the introduction of additional edits, as illustrated in the MSS/PSNR distortion plots, our method still ensures the precise preservation of the MSS.  From the MSS/PSNR distortion, we can observe that there is a certain balance between OBR and the preservation of MSS. When the original bitrate is lower, indicating a higher data compression ratio, this implies a higher error bound during compression. For a global topological descriptor like the MSS, even minor data alterations can lead to significant distortions in the MSS, potentially necessitating more edits. Conversely, when the original bitrate is higher, the deviation between the decompressed and original data is reduced, lowering the error rate in the MSS and decreasing the necessary edits.

\section{Performance Evaluation}

We evaluate the scalability and performance of our algorithm with both shared-memory CPU and GPU implementations. We use two datasets, Nyx ($512^3$) and viscous fingering ($128^3$), for performance studies because of their larger size and higher topological complexity than others. We use one CPU node (2$\times$ AMD EPYC 7763 with 512 GB of DDR4 memory) and one GPU node (single AMD EPYC 7763 CPU with 256 GB of DDR4 DRAM and four NVIDIA A100 GPUs) on the Perlmutter supercomputer at National Energy Research Scientific Computing (NERSC). Our implementation is based on C++, OpenMP, and CUDA; only a single GPU is used for benchmarking.

We measure performance with the end-to-end running time and four mini-benchmarks that evaluate the parallelization of (1) finding false critical points, (2) fixing critical points, (3) updating directions, and (4) MSS computation. For the mini-benchmarks, we artificially reset the data to the initial status the first time the subroutine is called. To ensure robustness and accuracy in our results, each mini-benchmark was executed 1,000 times, and the mean running time was recorded.

\subsection{Scalability on CPUs}

\begin{figure*}[!htb]
\centering
\includegraphics[width=\linewidth]{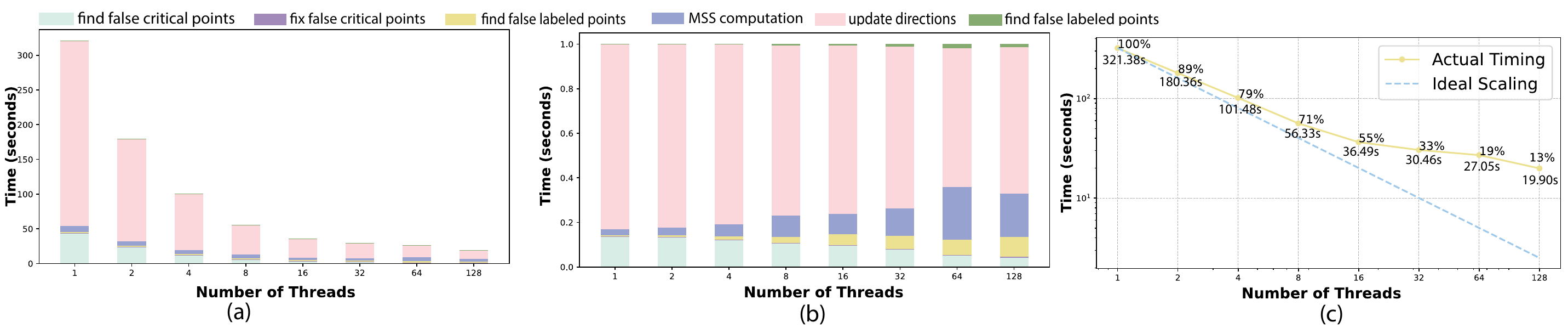}
\caption{Strong scalability study for the viscous fingering dataset with OpenMP: (a) performance breakdown and (b) percentage breakdown of computational tasks w.r.t the number of threads; (c) strong scalability study on the end-to-end time; percentages at data points represent the observed efficiency, and numbers show the actual timing.}
\label{fig:openmp_break}
\end{figure*}

While we recommend using GPUs for the practical use of our method, we study the scalability of our parallel algorithms on CPUs with different amounts of resources.
Figure~\ref{fig:openmp_break} shows the decreasing end-to-end running time with increasing CPU threads. We also demonstrated the timing breakdown for individual subroutines; for example, updating directions takes up to 80\% of the time because all vertices' directions must be updated whenever there is a change in the data. 

\begin{figure*}[!htb]
\centering
\includegraphics[width=\linewidth]{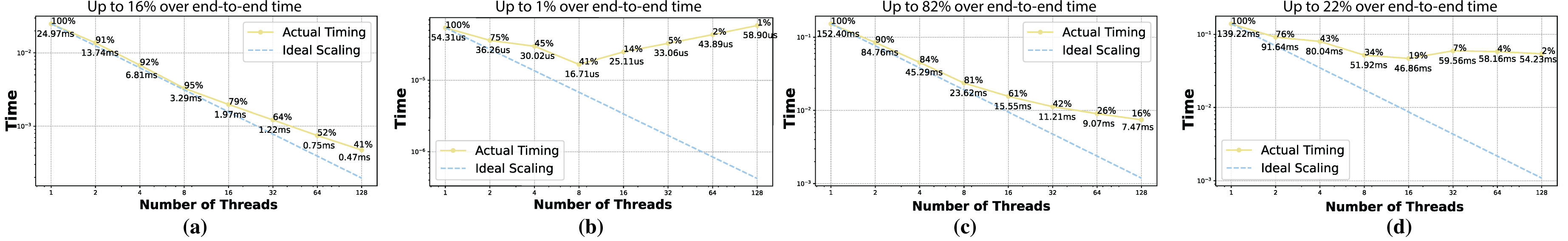}
\caption{Strong scalability study of the four tasks w.r.t the number of threads used in our experiment (blue line: actual timing, red line: ideal scaling), percentages at data points represent the observed efficiency, and numbers show the actual timing. (a) find the false critical points; (b) fix false critical points; (c) update directions; (d) MSS computation.}\label{fig:openmp}
\end{figure*}

Figure~\ref{fig:openmp} demonstrates the timings of the four mini-benchmarks with different numbers of threads. 
Note that the mini-benchmarks do not equally scale to 128 cores because of (1) the number of parallel tasks, (2) load balancing, and (3) locality and contention.  
First, the number of parallel tasks vary in different components of our algorithm.  For example, the number of (false) critical points is usually much smaller than the number of vertices; the former usually leads to insufficient utilization of hardware resources, thereby limiting scalability (Figure~\ref{fig:openmp}(b)).  
Second, in the MS segmentation computation (Figure~\ref{fig:openmp}(d)), we follow Maack et al.~\cite{MSS} to terminate an integral line and reuse existing integral lines; this strategy reduces computational cost but leads to imbalanced workloads.  
Third, our CPU implementation does not consider non-uniform memory access (NUMA) architectures, while our compute node has 8 NUMA domains; meanwhile, concurrent read/write also cause memory contentions even with lock-free data structures.  Besides, the underlying operating system may interfere with our parallel execution with many threads.  That said, the overall workflow remains scalable as we continue adding threads because the less efficient components (fixing critical points and computing segmentations) only account for up to 23\% of the end-to-end time.

\subsection{Performance on GPUs}

\begin{table}[ht!]
\centering
\caption{Timings of various tasks of our method on the Nyx dataset.} 
\setlength{\tabcolsep}{2pt}
{\small
\begin{tabular}{l|ccccc}
& Serial & \begin{tabular}{c}OpenMP\\(Optimal)\end{tabular} & CUDA & \begin{tabular}{c}GPU Accel.\\cf. OpenMP\end{tabular} & \begin{tabular}{c}GPU Accel.\\cf. Serial\end{tabular} \\ 
\hline
\textbf{End-to-end} &1192.38s & 110s & 6.31s & 17$\times$ & 198$\times$\\ 
Find false critical points & 1.202s & 0.26s & 0.008s & 32$\times$ & 150$\times$\\ 
Update directions & 10.46s & 0.6s & 0.014s & 42$\times$ & 700$\times$\\ 
Fix false critical points & 0.009s & 0.002s & 0.00038s & 5$\times$ & 23$\times$\\ 
MSS computation & 17.87s & 2.7s & 0.051s & 52$\times$ & 350$\times$\\ 
\end{tabular}
}
\label{tab:accx}
\end{table}
Table~\ref{tab:accx} shows GPU acceleration of the end-to-end and mini-benchmarks. We achieved a $17\times$ acceleration in the end-to-end performance compared with using all 128 threads on the CPU node; the acceleration is $198\times$ compared with the serial execution. 
For the mini-benchmarks, compared with serial execution with one single CPU, our GPU implementation achieves up to $700\times$ speed up across all tasks. 

\begin{table*}[ht]
\centering 
\scriptsize
\resizebox{\textwidth}{!}{
\begin{tabular}{c|c|ccc|ccc|ccc|ccc|cc|cc}

Dataset     
& Dimensions              
& \multicolumn{3}{c|}{Ours-SZ ($10^{-6}$)} 
& \multicolumn{3}{c|}{Ours-ZFP ($10^{-6}$)} 
& \multicolumn{3}{c|}{Ours-SZ ($5 \times 10^{-6}$)} 
& \multicolumn{3}{c|}{Ours-ZFP ($5 \times 10^{-6}$)}      
& \multicolumn{2}{c|}{GZIP} 
& \multicolumn{2}{c}{ZSTD}      \\ 
& &  $t_{comp}$ & $t_{fix}$ & OCR & $t_{comp}$ & $t_{fix}$ & OCR
& $t_{comp}$ & $t_{fix}$& OCR & $t_{comp}$& $t_{fix}$ & OCR
& $t_{comp}$
& CR 
& $t_{comp}$
& CR \\ \hline

AT & $177 \times 95 \times 48$ 
& 0.19 & 0.35 & 3.69
& 0.12 & 0.35 & 3.04
& 0.17 & 0.35 & 5.36
& 0.13 & 0.28 & 5.04
& 0.24 & 1.07 
& 1.18 & 1.17 
\\
        
NYX & $512^3$
& 13.8 & 3.80 & 7.76
& 11.8 & 4.80  & 4.20
& 12.98 & 6.56 & 7.48
& 10.75 & 8.17 & 4.74
& 67.8 & 1.17 
& 9.25 & 2.03 \\
        
Heated Flow & $150 \times 450$
& 0.08 & 0.29 & 2.74
& 0.03 & 0.33 & 3.38
& 0.08 & 0.27 & 3.58
& 0.04 & 0.35 & 3.76
& 0.64 & 1.91 
& 1.21 & 2.00 \\

Red Sea & $500 \times 500 \times 50$ 
& 0.98 & 0.51 & 52.7
& 0.42 & 0.51  & 31.4
& 0.88 & 0.55 & 62.5
& 0.39 & 0.46  & 34.13
& 4.79 & 26.4 
& 0.16 & 31.38 \\
        
CESM-ATM & $1800 \times 3600$
& 0.60 & 0.61 & 5.96
& 0.64 & 0.42 & 3.13
& 0.74 & 1.41 & 6.18
& 0.65 & 0.98 & 3.46
& 3.37 & 2.08 
& 1.65 & 2.27 \\

\end{tabular}}
\caption{Overhead in timings and compression ratios compared with lossy (SZ3 and ZFP) and lossless (GZIP and ZSTD) compressors across different datasets. $t_{comp}$ and $t_{fix}$, respectively, represents the timings (in seconds) of these compressors and our algorithm.}
\label{tab:dataset}
\end{table*}

\begin{figure*}[!htb]
    \centering
    \includegraphics[width=\textwidth]{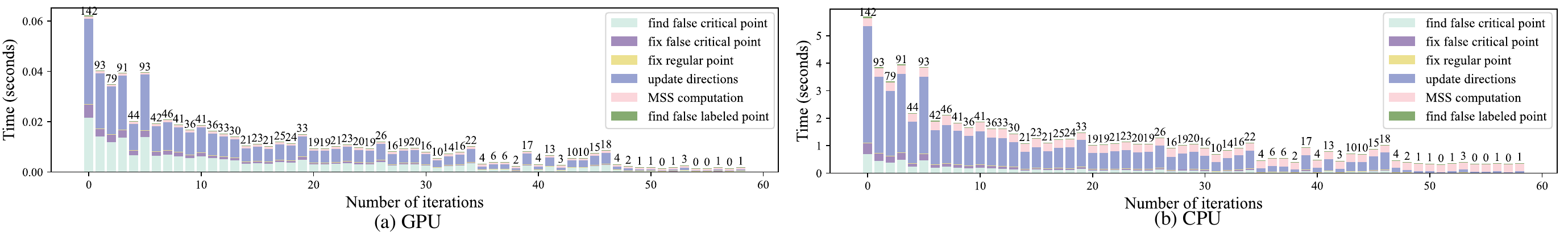}
    \caption{Timings of outer loop iterations on (a) GPU and (b) CPU with 32 threads; numbers above each bar chart denotes the number of sub-iterations.} 
    \label{fig:iteration}
\end{figure*}
Figure~\ref{fig:iteration} presents an additional performance visualization across multiple C- and R-loops, comparing performance on a CPU (with 32 threads) and a GPU. 
Across both settings, a notable decrease in time and the number of sub-iterations is observed with increasing iteration. This is because the number of false critical/regular points decreases as the iteration progresses. In each iteration, updating directions for each vertex accounts for a significant portion of the time, approximately 80\% for CPU and 50\% for GPU. This is followed by the process to find false critical points, around 10\% for CPU and 40\% for GPU.
Notice that measuring the impacts of atomic operations for concurrent read/write on GPU is an open challenge; we leave a formal evaluation of hardware utilization, throughput, and power consumption for the future.

\section{Discussion and Limitation}

\textbf{Computation and storage overhead.} Preserving MS segmentations comes with a cost, and users may need additional studies to balance their applications' data reduction, performance, and feature preservation needs.  
Table~\ref{tab:dataset} exemplifies the overhead in timings and compression ratios with multiple datasets.  We also compare performance with GZIP and ZSTD, which losslessly compress data and preserve MS segmentations accurately.
Our method incurs costs to store the edits as additional metadata attached to the native lossy compression outputs, but still delivers reasonable overall compression ratio with feature preservation.  
Our overall compression ratios are also better than lossless compressors, which only deliver $\sim2\times$ ratios for most floating-point data~\cite{zfp} (except in cases with empty spaces like the Red Sea data).
With an error bound of $10^{-6}$ based on SZ3, our overall compression ratio for the $512^3$ Nyx data is $7.76\times$, while GZIP is $1.17\times$.  
On the computation overhead, our timings are within the same order as the time to compress data with SZ/ZFP. 
For example, SZ3 takes $\SI{13.8}{s}$ to compress the Nyx data with $10^{-6}$ error bound, and our algorithm takes an additional $\SI{3.8}{s}$ to correct MS segmentations, while GZIP takes $\SI{62.8}{s}$.  
Although the computation and storage overheads vary across different datasets with different dimensionalities and error bounds, we justify the extra time cost considering the feature-preservation capabilities essential for deriving scientific insights.

\textbf{Preservation of saddles.} First, preserving saddles is unnecessary for preserving PLMSS because saddles are not constituents of PLMSS.  For example, in the AR application in Figure~\ref{fig:case_study}(a), the original and edited data have identical MS segmentations, yet two false saddles exist in the latter. Erroneous saddles do not affect this application because the characterization of ARs relies solely on the segmentation boundaries.  Second, as reviewed in Section~\ref{sec:background} and by Maack et al.~\cite{MSS}, PLMSS cannot capture saddle-saddle pairs and does not support downstream tasks that require the full MS complex; in other words, PLMSS's limitations apply to our method.  Third, it is possible to generalize our method in the future to preserve saddles with additional computation and storage costs.  Preserving a saddle requires maintaining value ordering among multiple neighboring vertices, making it nontrivial to design an iterative strategy to guarantee convergence.

\section{Conclusions and Future Work}
We introduce a novel method for preserving PLMSS in lossy decompressed data. 
Our strategy generates a set of edits at the time of compression. 
These edits are then applied to the decompressed data, ensuring precise MSS reconstruction while preserving the error bound. 
Our approach also incorporates a parallel implementation that substantially accelerates our algorithm. 
We evaluate our methods with datasets from molecular dynamics, climate, combustion, and cosmology applications. 

We plan to improve our method in various aspects. 
First, the compression of edits is achieved through lossless compression, yet this aspect offers substantial room for improvement, potentially enhancing the final compression ratio. 
Second, our method can be extended to preserve the full MS complex by incorporating the preservation of saddles and saddle-saddle connectors.

\acknowledgments{This research is supported by the U.S. Department of Energy, Office of Advanced Scientific Computing Research (DE-SC0022753) and the National Science Foundation (OAC-2311878, OAC-2313123, OAC-2313124, IIS-1955764, OAC-2330367, OAC-2313122, and OAC-2327266). This research used resources of the National Energy Research Scientific Computing Center (NERSC), a Department of Energy Office of Science User Facility.}

\bibliographystyle{abbrv-doi-hyperref-narrow}

\bibliography{refs-msz}

\clearpage
\appendix
\section{DETAILS OF THE DATASETS}

The \textbf{Red Sea} dataset is a time-varying three-dimensional flow field and scalar field that changes over the regular grid. We have taken the scalar field from one time step for our experiments. This dataset is from the IEEE Scientific Visualization Contest 2020~\cite{SciVis2020}.

The \textbf{Adenine Thymine} dataset represents a molecular simulation detailing electron density confined to a plane within a three-dimensional space. This dataset is from the TTK Tutorial Data~\cite{ttk2020data}.

The \textbf{Viscous Fingering} dataset consists of an ensemble simulation that captures the viscous finger inside a cylindrical container filled with pure water and topped with an infinite amount of salt. The viscous finger represents regions of high concentration and occurs due to diffusion processes. Their formation is because salt is denser than water, causing the denser liquid to collect towards the lower half of the cylinder. This dataset is from the IEEE Scientific Visualization Contest 2016~\cite{SciVis2016}.

The \textbf{Nyx} dataset~\cite{Nyx} is a series of simulation datasets generated from the cosmological hydrodynamics simulation code Nyx~\cite{Almgren_2013}. It is a three-dimensional spatial array of data for post-analysis, including dark matter density and temperature.

The \textbf{Earthquake} dataset~\cite{Nyx} is an ensemble dataset representing a physical simulation of a magnitude 7.7 earthquake south of the San Andreas Fault. The dataset was obtained using a finite-difference code on the DataStar computer at the San Diego Supercomputer Center (SDSC). This dataset is from the IEEE Scientific Visualization Contest 2006~\cite{SciVis2006}.

The \textbf{Heated Flow} dataset is an ensemble simulation of two-dimensional flow generated around a heated cylinder, produced using the Gerris flow solver. In our experiments, we utilized a single time step from the data. This dataset is from the Computer Graphics Laboratory~\cite{heated}.

The \textbf{CESM-ATM} dataset is an ensemble simulation of atmospheric conditions using the Community Earth System Model (CESM)~\cite{cesm} with 60 time steps; we used one of the time steps in our experiment. The dataset is available at Scientific Data Reduction Benchmarks~\cite{sdrbench}.

The \textbf{Integrated Vapor Transport} dataset provides an ensemble of hourly integrated vapor transport (IVT) data from the MERRA-2 reanalysis. This data indicates the amount of water vapor transported over a single grid per second. The dataset can be accessed through the Earth System Grid Federation portal~\cite{ivt}.

The \textbf{Combustion} dataset is an ensemble simulation the intricate dynamics of turbulent combustion processes. These simulations provide detailed insights into flames' behavior, fuel and oxidizer mixing, heat release, and turbulence-chemistry interactions. The dataset is made available by Jackie Chen at Sandia National Lab through the SciDAC Institute~\cite{combustion}.

\section{EVALUATION METRICS}
We review the evaluation metrics used in our paper.
In this context, $N$ represents the number of data points.

\textbf{PSNR distortion} refers to the trade-off between the bitrate and PSNR. It describes how the bitrate affects the PSNR of the decompressed data, where a higher bitrate generally results in a higher PSNR, indicating better quality. This is typically represented as a curve with the x-axis showing the bitrate and the y-axis showing the PSNR of the decompressed data.

Assuming $f$ and $f'$ represent the original and decompressed scalar values, respectively, the Peak Signal-to-Noise Ratio (PSNR) is calculated using the following equation:
\begin{equation}
    \text{PSNR} = 20 \cdot \log_{10} \left( \frac{\text{MAX}_f}{\sqrt{\text{MSE}}} \right)
\end{equation}
 
where $\text{MAX}_f$ is the maximum possible value of the original data $f$, and $\text{MSE}$ is the Mean Squared Error between the original and decompressed data, defined as:
\begin{equation}
\text{MSE} = \frac{1}{N} \sum_{i=1}^{N} (f_i - f'_i)^2
\end{equation}

\textbf{Right labeled ratio} is the percentage of points in the data with correct MSS labels. Assume that $N_f$ is the number of data points with the wrong MSS label; the right labeled ratio is calculated by the following equation: 
\begin{equation}
    \text{Right labeled ratio} = 1 - \frac{N_f}{N}
\end{equation}

\textbf{MSS distortion} refers to the trade-off between the bitrate and Right labeled ratio. It is similar to PSNR distortion, except that the y-axis is replaced with the right labeled ratio.
\section{ADDITIONAL EXPERIMENTS}
\textbf{Discussion on error distribution}
\begin{figure}[!htp]
    \centering
    \includegraphics[width=1.0\columnwidth]{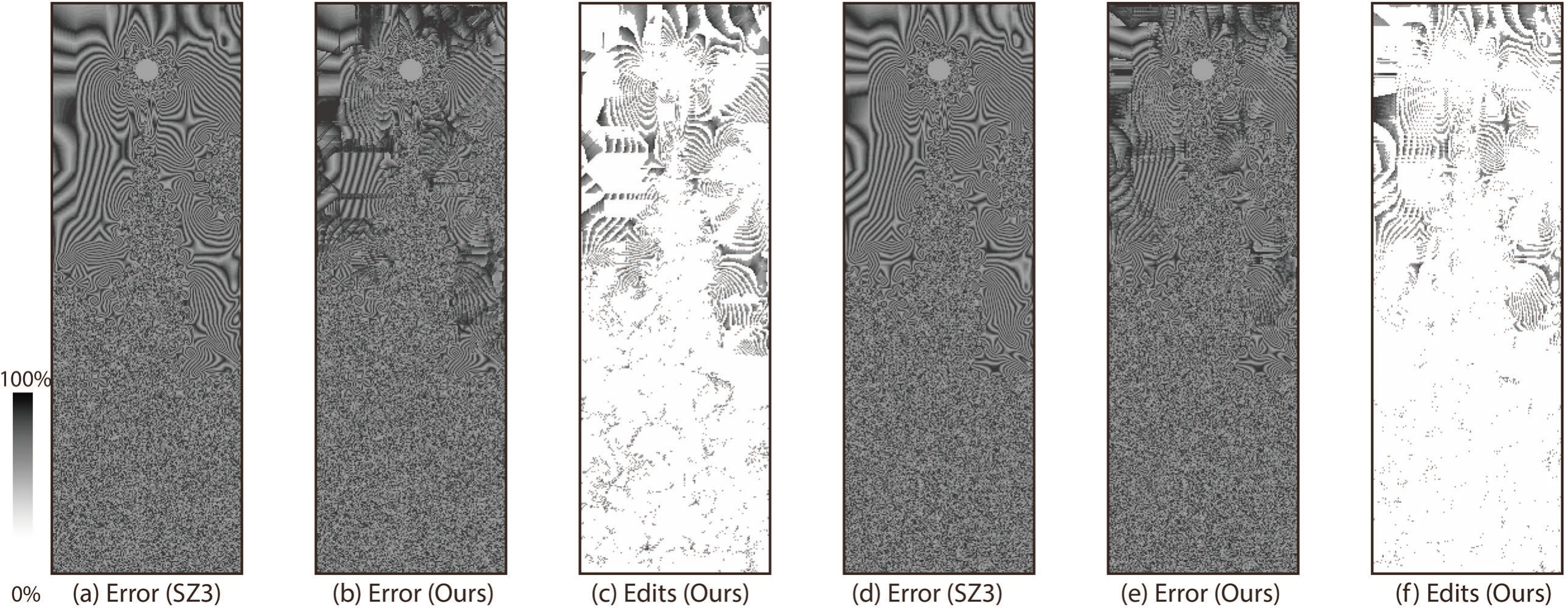}
    \caption{Spatial distribution of error and edits with error bounds of $7 \times 10^{-4}$ (a, b, and c) and $10^{-3}$ (d, e, and f) based on SZ3.}
    \label{fig:error distribuction}
\end{figure}
We visualized the error map between the decompressed data from SZ3/Ours-SZ3 and the original data, as shown in Figure~\ref{fig:error distribuction} (a), (b), (d), and (e). Additionally, we visualize the edits applied by our method, as illustrated in (c) and (f). Our edits are sparsely distributed yet form continuous patches within the domain. This observation justifies our approach of sorting the edit indices in ascending order and then compressing the differential sequence.

\end{document}